\documentclass[aps,preprint]{revtex4-1}
\usepackage{graphicx}
\usepackage{amsmath}
\usepackage{makeidx}
\usepackage{amsfonts}
\usepackage{amssymb}
\usepackage{dsfont}
\usepackage{color,soul}

\usepackage{epstopdf}

\begin{document}

\title{Circuit topology for bottom-up engineering of molecular knots}

\author{Anatoly Golovnev$^{a}$, Alireza Mashaghi$^{a,\ast}$}
\affiliation{$^a$Medical Systems Biophysics and Bioengineering, Leiden Academic Centre for Drug Research, Faculty of Science, Leiden University, 2333CC Leiden, The Netherlands.}
\affiliation{$\ast$ a.mashaghi.tabari@lacdr.leidenuniv.nl}

\date{\today}

\begin{abstract}
The art of tying knots is exploited in nature and occurs in multiple applications ranging from being an essential part of scouting programs to engineering molecular knots. Biomolecular knots, such as knotted proteins, bear various cellular functions and their entanglement is believed to provide them with thermal and kinetic stability. Yet, little is known about the design principles of naturally evolved molecular knots. Intra-chain contacts and chain entanglement contribute to folding of knotted proteins. Circuit topology, a theory that describes intra-chain contacts, was recently generalized to account for chain entanglement. This generalization is unique to circuit topology and not motivated by other theories. In this paper, we systematically analyze the circuit topology approach to a description of linear chain entanglement. We utilize a bottom-up approach, i.e., we express entanglement by a set of 4 fundamental structural units subjected to 3 (or 5) binary topological operations. All knots found in proteins form a well-defined, distinct group which naturally appears if expressed in terms of these basic structural units. Prime knots, which are viewed by knot theory as undecomposable, are also made of these structural units connected in some specific way. In turn, this kind of connection shows the fundamental reason why prime knots cannot be decomposed in the rigorous sense of knot theory. We believe that such a detailed, bottom-up understanding of the structure of molecular knots should be beneficial for molecular engineering.
\end{abstract}

\maketitle

\begin{figure}[ht]
\centering
\includegraphics[angle=0, width=0.6\textwidth]{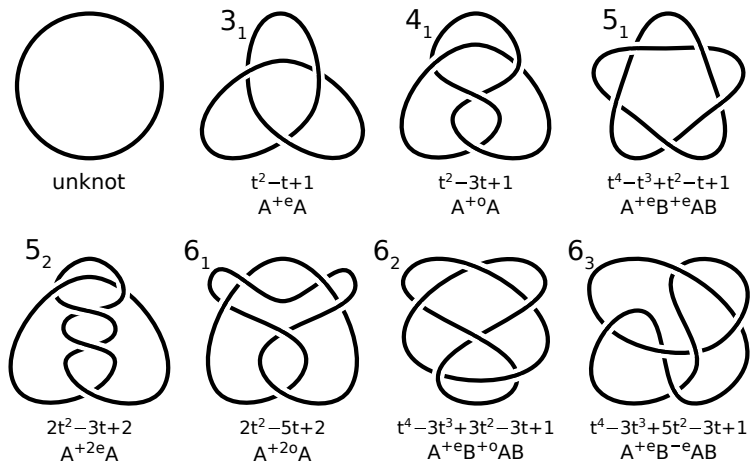}
\caption{Prime knots with $6$ and fewer crossings. Knot theory notation, Alexander polynomial, and circuit topology string notation are provided for each knot. $3_1$ is called a trefoil. $4_1$ is called a figure 8 knot. $3_1$, $4_1$, $5_2$, $6_1$ are the knots found in proteins.}
\label{prime_knots}
\end{figure}

Linear chains, such as proteins and nucleic acids, demonstrate an immense structural diversity owing, in part, to a myriad of possible chain configurations which appear as various knots\cite{lim}, slip-knots\cite{stasiak}, and loops, and are believed to be of relevance to biological function of these molecules \cite{biofunction, diseases}. A three-dimensional structure of linear molecular chains is commonly described in terms of knot theory \cite{adams}, which is a powerful and rigorous mathematical concept. The approach is generic and applicable to any linear chain, not limited to biological molecules. In terms of knot theory, a knot is a one-dimensional topological circle embedded into three-dimensional space; it is a continuous structure without free ends. In other words, in order to turn a linear chain into a knot, one has to join the chain ends. While discussing chains and knots, it might be convenient to think of a rope which we will tie and tangle. The most basic, ``undecomposable'' knots are called prime knots. Some of them are shown in Fig.\ref{prime_knots}, where the capital number specifies the number of crossings in the minimal crossing projection and the subscript is assigned in order to distinguish between knots with the same number of crossings. Here, the number of crossings in each knot cannot be decreased but can easily be increased by, for example, twisting some loops or threading the rope through a loop. Knots do not change upon deformations which do not break the rope, i.e., which do not break the continuity of the knot. Such deformations can be expressed via a sequence of specific deformations performed on a knot projection which are called Reidemeister moves. The resulting structure could look very different form the original primary knot, but is perceived as equivalent by knot theory. This is one of the major ideological differences between knot theory and molecular engineering. Knot theory is designed for other purposes, namely to capture topological invariance under ambient isotopy (i.e. weather two knots can be deformed into each other); while in case of molecular engineering, even minor changes in shape of the chain might matter. For example, slip-knots - which are very common in proteins and crucially important for their proper functioning - are ignored by knot theory. However, the nature of slip-knots is rather geometric than topological and therefore knot theory has to ignore them. In our work, we aim to develop a theory which would serve for molecular engineering. All basic molecular engineering operations should have a clear and intuitive representations. Hence, having the basic structural units to build up a chain seems to be convenient. For example, it is well known that by cutting the loop of a slip-knots, the chain is reduced to a trefoil. Our theory should (and will) give a clear {\it analytical} visualization of this process.

In this paper, we will consider chains with different geometrical shapes. In order to make sure that two chains cannot be deformed into each other, we will join their ends to form a mathematical knot and calculate a so-called Alexander polynomial. Its definition and calculation can be found in any knot theory textbook or in our previous paper \cite{ct}. If two knots have different Alexander polynomials, then these knots must be different. The inverse statement is usually correct, but not always.

An application of knot theory to proteins is centered around a search for prime knots in a spatial protein structure. The only knots which have been so far found in proteins \cite{flapan} are $3_1$, $4_1$, $5_2$, $6_1$. What is a fundamental reason for this choice? What property exactly separates these knots from other knots? The answer to the second question is known. The knots found in proteins can be formed following the so-called twisted hairpin folding mechanism \cite{taylor} outlined below. This mechanism is rather a phenomenological explanation which does not provide a fundamental difference between knots in terms of knot theory. In principle, a topological theory alone is not able to provide such a reason because physical properties of the chain must matter. In our theory, the twisted hairpin mechanism appears naturally as part of the formalism. A few years ago, the concept of circuit topology was suggested in order to account for intra-chain contacts \cite{2014,trends, notice}, which are also very important for proteins. Very recently, circuit topology was generalized to account for chain entanglement as well, focusing on applicability to real-life molecules \cite{ct}. The new framework is still in its development stage and lacks certain rigorousness, especially in comparison to very-well developed knot theory. In this study we attempt to strengthen the foundation of generalized circuit topology and demonstrate that this theory appeals to the ``natural and inherent'' language describing entanglement. To demonstrate this, we will, among others, re-discover some known results which appear smoothly as internal part of circuit topology. We believe it will be useful for molecular engineering and will help puzzle out the design principle of naturally evolved protein knots.  

\section{S-contacts}

\begin{figure}[ht]
\centering
\includegraphics[angle=0, width=0.6\textwidth]{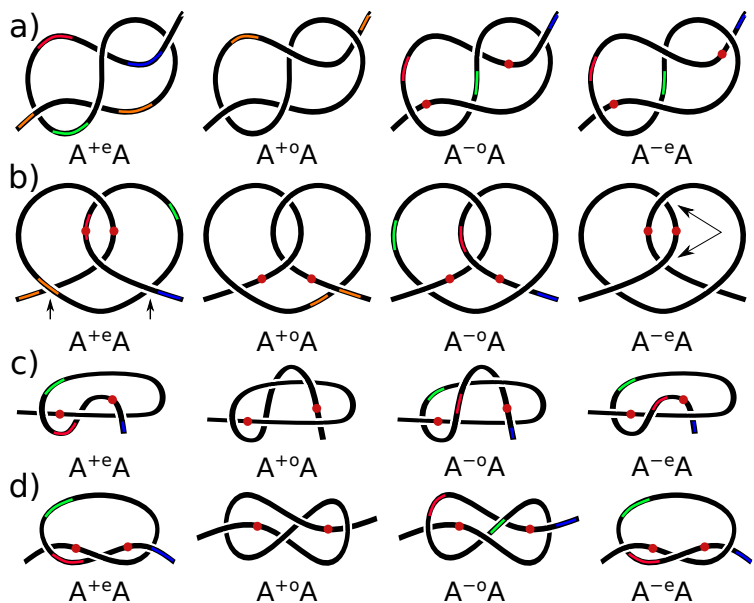}
\caption{A list of s-contacts. All the chains in each column are equivalent, i.e., can be deformed into each other. All the chains are 3D structures, not projections. a) The definition of s-contacts. b) The ``flat'' representation of s-contacts. c and d) Other equivalent representations of s-contacts where the loops are not easy to spot. A change of chirality, i.e., of the sign +/- in the notation, means flipping of all crossings in the representation. Red balls indicate contact sites. Colored stripes are added to make the equivalence of different representations more visual.}
\label{contacts}
\end{figure}

We are looking for basic structural units which would comprise a molecular knot and be invariant to the knot structure, i.e., would not change upon a knot deformation. By a deformation we understand any manipulations with a chain as long as the chain is not broken and its ends are kept away from the manipulations, so that we do not change the topological structure of the chain. Also, we recognize the fact that molecules are 3-dimentional, hence our formalism is built in 3D, so for the most part, we do not consider projections. On the other hand, it makes it harder to describe the mutual position of different segments of a chain. Unlike in 2D, in 3D there are no crossings of a chain and the corresponding segments of a chain might be distant in space, hence one cannot strictly define a loop. In what follows, these terms should be understood merely as references to certain segments or configurations of a chain in 3D, and as a tribute to the fact that all drawings are inevitably flat, i.e., 2D. A single loop, i.e., one twist of a chain (or a rope) which leads to only one crossing (or no crossings at all in certain projections), is not stable. It can easily be undone (untied) by pulling the chain ends, and hence cannot serve as a structural unit. To make it stable, one should ``fix'' the loop by threading the chain through it. Fig.\ref{contacts}a shows all 4 possible resulting structures. By threading the loop, we have to go around one segment of the chain, thereby creating another loop. So, each of the 4 structures is symmetric: each consists of two identical loops hooked together. Following the chain in either direction, left-to-right or right-to-left, we see the same structure. Each structure is called an s-contact, or a soft contact, and cannot by untied by pulling the chain ends. The term ``contact'' was coined to stress similarity to intra-chain, non-entanglement contacts which can also be treated by circuit topology \cite{ct} but are not considered in the present paper. Each contact has two contact sites. (In case of intra-chain contacts, contact sites are the two chain segments which are linked together.) An s-contact is supposed to be contained between its contact sites, i.e., contact sites are the boundaries of the structural unit as viewed while moving along the chain. We define that contact sites are located where the chain passes through the loops, so that the entangled segment is located in-between. What happens to the chain outside to the structural unit (i.e. on either side from the contact sites) is irrelevant. Since s-contacts should be considered as 3-dimentional entities, the exact position of contact sites depends on many parameters\cite{ct}, e.g., on the knot tightness, but it always represents the knot structure. Also, contact sites can migrate along the chain during a chain deformation. This uncertainty is essential in order to be able to catch the fixed topological structure of a flexible, mobile chain. Contact sites are depicted with a red ball on several structures shown in Fig.\ref{contacts}.

Contacts should be given names. We usually use capital letters, such as contact A, contact B, etc. In Fig.\ref{contacts}a, each chain has one contact. If we move along the chain from any end and write down contact sites we encounter, we will get AA, which is a code for one contact. How to distinguish between the four different s-contact shown in the figure? Fig.\ref{contacts}b shows the same s-contact from Fig.\ref{contacts}a in a different representation where loops are easier to spot. The both representations are topologically equivalent and one can continuously transition from one to another without breaking the chain. Each s-contact is a connection of 2 loops. Each connection has 4 crossings, 2 of which are independent. Each independent crossing can take two values, depending on which chain segment is on top. It means that there are only 4 s-contacts possible and hence our set of s-contacts is complete. The pair of crossings defining the s-contact chirality is depicted on the left-most s-contact. The chirality of each loop is defined by the conventional right-hand rule. If the loops have different chirality, then none of these loops has been fixed by threading the chain through it, which means this structure will untie. In other words, despite consisting of two loop, s-contacts can only be either positive, A$^+$A, or negative, A$^-$A.

The other independent pair of crossings, shown on the right-most s-contact in Fig.\ref{contacts}b, defines how the two loops are hooked together. If the chain passes through the loop in the same direction as the chain shifts in each loop at the crossings defining chirality, such an s-contact is called ``even'', A$^e$A. If the chain passes through the loop in the opposite direction, the s-contact is called ``odd'', A$^o$A. Hence, the only 4 possible s-contacts are A$^{+e}$A, A$^{+o}$A, A$^{-o}$A, A$^{-e}$A. This notation is called the string notation of circuit topology. It codes a chain entanglement as a string of letters. One of the advantages of this notation is the ability to apply combinatorial analysis directly to a description of entanglement. Also, note that the attributes introduced above are universal and do not depend on the chain orientation, i.e., in which direction we move along the chain, left-to-right or right-to-left.

\begin{figure}[ht]
\centering
\includegraphics[angle=0, width=0.6\textwidth]{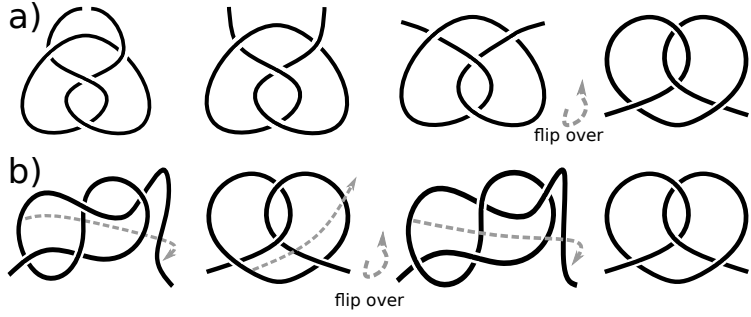}
\caption{Properties of s-contact A$^o$A. a) A$^o$A corresponds to $4_1$ knot (see Fig.\ref{prime_knots}). b) Transition between A$^{+o}$A and A$^{-o}$A, demonstrating that A$^{+o}$A$=$A$^{-o}$A$\equiv$A$^o$A.}
\label{achiral}
\end{figure}

Fig.\ref{contacts}c shows two other representations of s-contacts. Despite looking different, the representations from Fig.\ref{contacts}a and Fig.\ref{contacts}c are actually not that much distinct in 3D and can be transformed into each other with a slight deformation. We show them all in order to simplify visualization and to stress that everything should be viewed in 3D. Also note that it is hard to single out separate loops in Fig.\ref{contacts}c. The representation in Fig.\ref{contacts}b is topologically identical to the other representations but is more distinct from them. A transition to it requires a major deformation. This representation can for convenience be considered flat as a limiting case of 3D, which will be useful in the next sections. This ``flat'' representation resembles projections used by knot theory and can be useful in building a link between knot theory and circuit topology.

So far, we identified 4 stable ``basic units'' of chain entanglement, and called them s-contacts. It should make sense on the intuitive level because any messy blob of a rope is held together by loops hooking to each other, which is the essence of entanglement. Also note that if we flip only one crossing in any chain from Fig.\ref{contacts}, the s-contact will disappear and that chain will untie. Let us consider s-contacts in the view of knot theory. To form a knot from a rope, one has to join its ends. To make a rope form a knot, one has to cut the knot somewhere. A$^e$A corresponds to $3_1$. This knot can be right-handed, as in Fig.\ref{prime_knots}, or left-handed if all the crossings are flipped. A$^o$A corresponds to $4_1$, see Fig.\ref{achiral}a for a visualization of the sequence of corresponding deformations. This knot is known to be achiral (amphichiral), i.e., $+4_1=-4_1$, i.e., A$^{+o}$A and A$^{-o}$A can be deformed into each other. The sequence of corresponding moves is shown in Fig.\ref{achiral}b. However, both A$^{+o}$A and A$^{-o}$A should be kept and considered as separate s-contacts because, as will be shown below, they comprise different knots in a presence of other s-contacts. Can we distinguish A$^{+o}$A and A$^{-o}$A when they are along? Topologically speaking, we cannot, and knot theory is clear about this. Geometrically speaking, it boils down to the notion of stability, similar to retaining or quenching slip-knots. Namely, the deformation needed for a transition between A$^{+o}$A and A$^{-o}$A costs energy. If the cost is low, the transition can occur spontaneously. Otherwise, it will not occur, rendering the molecular knot stable. We will discuss it in Section 3.2. However note in Fig.\ref{contacts}d that A$^{+o}$A and A$^{-o}$A look like a mirror reflection. Such a flip of symmetry matters in proteins, so we must retain it for molecular engineering purposes.

We claim that any knot can be considered as a set of only 4 s-contacts connected according to the rules discussed in the next sections. However, s-contacts might not be easy to spot, even in such a simple case of $4_1$ knot. Where exactly are s-contacts in the prime knots shown in Fig.\ref{prime_knots}? We will break it down below, but at the current state of our theory, it is easier to go in the opposite direction, i.e., to tie s-contacts on a rope and then identify the resulting knot.

\section{SPX configurations of s-contacts}

One s-contact has two contact sites and appears in the string notation as AA where each letter signifies this contact's sites. Two s-contacts, AA and BB, can occur in three different configurations defined by permutations of two pairs of letters: AABB, ABBA, ABAB. Because s-contacts can have any name, the configurations ABAB and BABA are identical. These three configurations are called series (S), parallel (P), and cross (X) and comprise the SPX relations. Regardless how many s-contacts a knot consists of, these pair-wise relations always hold and are shown unambiguously by the string notation. For example, in AACDCBDB all relations are immediately obvious, e.g., the contacts A and B are in series, and contacts C and D are in cross, etc. Therefore, a consideration of all pairs of s-contacts is sufficient to describe a chain entanglement. Indeed, the string notation lists all contact sites as they appear along the chain and can be unambiguously deduced from the relative positions of all pairs of contact; thereby it completely specifies (or codes) the chain entanglement in terms of s-contacts.

\begin{figure}[ht]
\centering
\includegraphics[angle=0, width=0.6\textwidth]{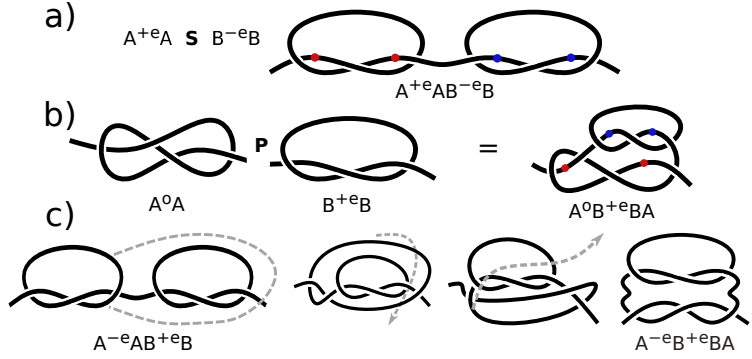}
\caption{SP relations. a) Two s-contacts in series. b) Two s-contacts in parallel. c) A transition between series and parallel configurations.}
\label{sp}
\end{figure}

Series and parallel configurations are easy to visualize. Fig.\ref{sp}a shows two s-contacts in series; Fig.\ref{sp}b shows two other s-contacts in parallel; and contact sites are marked with colored balls. Note that in Fig.\ref{sp}b the internal contact, marked with blue balls, can be placed in different loops of the external contact marked with red balls. However, blue balls will always be between red balls, i.e., the contacts are always in parallel with the same string notation (ABBA). Here we do not specify the symmetry and chirality because this rationale works for any kind of s-contact. SP configurations with other types of s-contacts can be drawn in a similar manner. Note that chains from Fig.\ref{sp}a and Fig.\ref{sp}b look very different not only because they consist of different s-contacts, but also because of the different relations between s-contacts, i.e. series and parallel. However, the relations of s-contacts can be swapped between each other by a sequence of moves shown in Fig.\ref{sp}c. Notice that contact B is not deform or altered in any way during the deformation. Hence, it can be replaced by any other kind of s-contact or any arrangement of s-contacts. Contact A throws out a loop and gets deformed. A similar deformation can be applied to any kind of s-contact. Applying this deformation to one s-contact after another, one can push s-contacts inside other s-contacts or pull them outside, thereby dragging an s-contact along the string in string notation. For example, one can turn A$^{e}$A\,B$^{o}$B\,C$^{e}$C into A$^{e}$A\,C$^{e}$C\,B$^{o}$B (i.e., $3_1\,4_1\,3_1$ to $3_1\,3_1\,4_1$). In general, because a pair-wise consideration is sufficient, any set of s-contacts consisted of only SP configurations can have any fraction of S and P relations as long as their total number is constant. As limiting cases, such a set can be deformed into all s-contacts in series or all s-contacts in parallel. Also, because s-contact can be deformed into each other, A$^{+o}$A and A$^{-o}$A are indistinguishable in case of series and parallel configurations. However, A$^{+e}$A and A$^{-e}$A cannot be deformed into other s-contacts. Two entangled chains consisted only of SP configurations can be deformed into each other only if they contain the same number of s-contacts of each kind. For example, the chains in SP configurations from Fig.\ref{sp}a and b contain different kinds of s-contacts and therefore cannot be deformed into each other.

The transition from Fig.\ref{sp}c is important in the context of protein folding and has been studied in the literature \cite{sptran}. Knot theory cannot distinguish these two configurations because they correspond to the same knot\cite{ct}. However, in real molecules such a transition between these configurations requires energy and once again comes down to the question of stability; the transition might be very probable or might never happen, depending on the physical properties of the chain. Circuit topology aims at addressing this question and provides consideration of different levels of structural stability.

SP configurations are similar to the notion of a connected sum used in knot theory. To form a connected sum of two knots, one should cut each knot and merge together the resulting ends, which is the procedure demonstrated in Fig.\ref{sp}. Consequently, based on the connected sum properties, the Alexander polynomial of several s-contacts in series or in parallel is a product of Alexander polynomials of each s-contact.

\begin{figure}[ht]
\centering
\includegraphics[angle=0, width=0.6\textwidth]{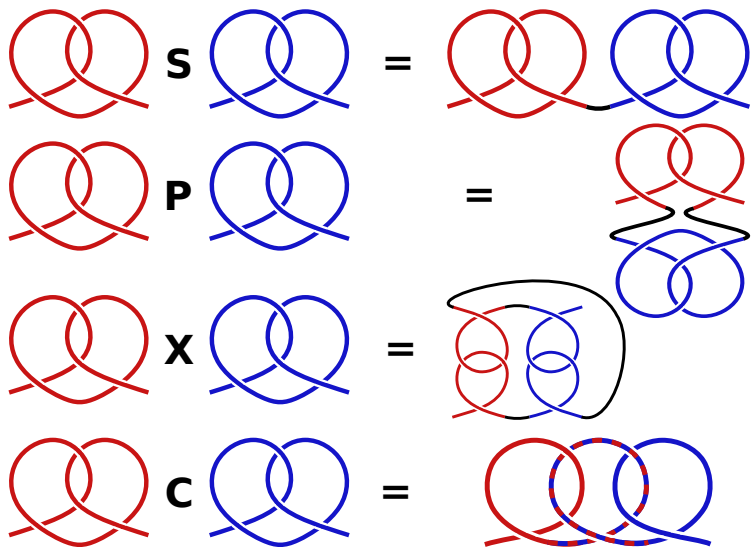}
\caption{SPX and concerted (C) configurations of two s-contacts in the loop representation.}
\label{loops}
\end{figure}

Cross (X) configurations are very different from SP configurations. Each s-contact consists of two loops. Fig.\ref{loops} shows all possible arrangements of two pairs of loops. This ``flat'' representation is convenient for listing and counting cases, but one should keep in mind that these flat structures can always be deformed into 3D. In S and P configurations, the loops belonging to the same s-contacts are hooked together. Each pair of loops can be separated and cut off the whole chain. It is not possible to do in X configuration where loops belonging to different s-contacts are connected. Contacts A$^{+o}$A and A$^{-o}$A are identical, i.e., can be deformed into each other, only as long as their loops are free to move as is the case in SP configurations, see Fig.\ref{achiral}. In X configuration, loops from one s-contact are connected to loops from another s-contact, and hence are not free. Therefore, contacts A$^{+o}$A and A$^{-o}$A are not identical and lead to different knots when they are parts of X configuration.

Let us consider the transition between S and P configurations by looking at the illustrations in Fig.\ref{loops} and see again that it does not work for X configuration (C configuration will be considered in the next section). Also, the single deformation shown in Fig.\ref{sp}c is obvious only in case of two s-contacts. What if there are more s-contacts? What is the general rule? Topologically, we can stretch the chain but we should not break it. When two loops are joined into an s-contact, they are connected and cannot be separated. In contrast to loops, one single s-contact can be moved along the chain freely. So, we take the blue contact from the right top corner of Fig.\ref{loops} and move it to the left. It passes through the left loop of the red contact and then moves down to the bottom of the red contact. By this manipulation we turned S configuration into P. In other words, we are only allowed to move the whole contact along the chain, but not contact sites. The transition AABB $\longrightarrow$ ABBA should be understood as the whole contact B moves to the left, but not as one contact site of contact A moves to the right. Following the same logic, let us add contact C to the left of X configuration in Fig.\ref{loops}. The resulting string is CC\,ABAB. Contact C as one entity can move through the loop of contact A to form A\,CC\,BAB or further to ABCCAB; but it can never lead to CACBAB or ACBCAB. As a consequence, X configuration cannot be turned into SP. Here, the crossings in Fig.\ref{loops} make no difference in the rationale, hence this conclusion is applicable to any kind of s-contacts.

\begin{figure}[ht]
\centering
\includegraphics[angle=0, width=0.6\textwidth]{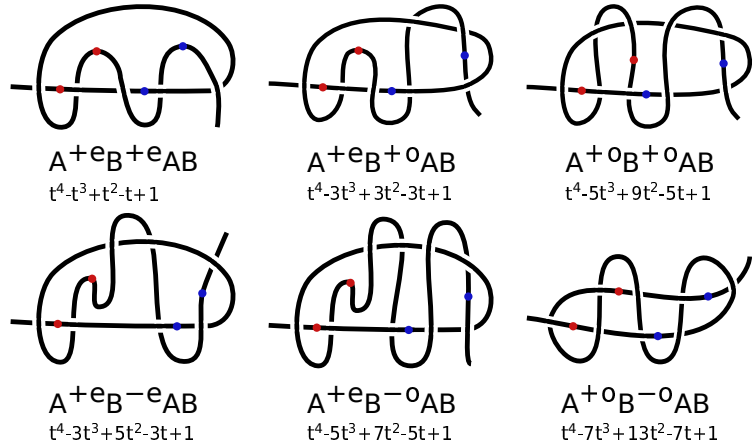}
\caption{All independent (up to symmetry) configurations of two s-contacts in cross. Contact sites are shown as colored balls. String notation and Alexander polynomials are provided.}
\label{x}
\end{figure}

There are 4 s-contacts (A$^{+e}$A, A$^{-e}$A, A$^{+o}$A, A$^{-o}$A), which gives rise to $4^{2}=16$ configurations of 2 s-contact in cross, which can be written as a 4x4 table. This table is symmetric owning to the left-right symmetry, i.e., which direction we move along the rope, e.g., A$^{+e}$B$^{-e}$AB vs. B$^{-e}$A$^{+e}$BA. This leaves 4 configurations on the diagonal and 6 configurations above (or below) the diagonal, i.e., 4+6=10 configurations. As shown in Fig.\ref{contacts}, a change of chirality in a single s-contact, i.e., a change of all signs in the string notation, leads to a flipping of all crossings. This property holds for any combination of s-contacts. Indeed, flipping {\bf all} crossings means a mirror reflection of the whole knot. Due to this symmetry, the number of configurations to consider can be further reduced. On the diagonal, A$^{+e}$B$^{+e}$AB and A$^{-e}$B$^{-e}$AB can be treated (or drawn) as one configuration. Similarly, A$^{+o}$B$^{+o}$AB and A$^{-o}$B$^{-o}$AB. The same holds for two non-diagonal configurations: A$^{+e}$B$^{+o}$AB vs. A$^{-e}$B$^{-o}$AB and A$^{+e}$B$^{-o}$AB vs. A$^{-e}$B$^{+o}$AB. It leaves us with 10-4=6 configurations which are shown in Fig.\ref{x}. Let us count these 6 configurations again, but this time geometrically. First, we tie contact A whose contact sites are marked with red balls. We made one of the loops corresponding to contact A the largest in the illustration in order to make it easier to spot contact A. This loop can always be shrunk without changing the overall topology. After contact A is ``fixed'', i.e., after the second red ball, where can the rope go? It can go away, which would create SP configurations. Or the rope can pass through this large loop again, thereby creating another s-contact in cross with contact A. Fig.\ref{x} shows all possible route of the rope leading to another s-contact. If the rope keeps passing though the large loop, it will just create more and more s-contacts. Note that A$^{+e}$B$^{+o}$AB and A$^{+e}$B$^{-o}$AB look very different and their Alexander polynomials are different, though the only difference between them is the chirality of contact B. It is a rigorous proof that even though B$^{+o}$B and B$^{-o}$B are identical while standing along, i.e., while being in series or in parallel to other s-contacts, in cross configurations they lead to different composite structures. Hence, all 4 s-contacts should be retained.

In order to easily verify and visualize the correctness of strings provided in Fig.\ref{x}, one can untie one of the s-contacts. For example, in A$^{+e}$B$^{+o}$AB one can unhook the left site of contact A (red balls), so that contact A disappears and only B$^{+o}$B is left. In the drawing, this procedure means that the left-most crossing is flipped, so that the left red ball no longer passes through the loop. Then the rope can be deformed into the configuration from Fig.\ref{contacts}c by reducing the loop freed by the flipping. The same holds for the right site of contact B (blue balls). Also, consider A$^{+e}$B$^{-o}$AB. After the second red ball, the rope wraps around the large loop and gets ``fixed'' by passing through the large loop at the location of the second blue ball. If the rope does not pass through the large loop, so there is no blue ball, then no matter how many times the rope wraps around the large loop, the second contact will not be formed. Indeed, this ``spiral'' around the large loop will not be stable and will be easily untied by pulling the rope ends apart.

Each of the four s-contacts has an Alexander polynomial degree 2. Two s-contacts in any of SPX configurations have Alexander polynomial degree 4. $n$ s-contacts in any SP configuration have Alexander polynomial degree $2n$, because it is a product of Alexander polynomials of single s-contacts. It is reasonable to expect that the Alexander polynomial degree scales the same for X configurations as well. Indeed, Fig.\ref{x} shows a clear pattern of Alexander polynomials, depending on the kind of s-contacts in cross. The easiest pattern appears for positive even contacts. A$^{+e}$A corresponds to $t^2-t+1$; A$^{+e}$B$^{+e}$AB corresponds to $t^4-t^3+t^2-t+1$. One can predict that A$^{+e}$B$^{+e}$C$^{+e}$ABC has the Alexander polynomial $t^6-t^5+t^4-t^3+t^2-t+1$. The corresponding prime knot, $7_1$, is not drawn here, but can be easily deduced from the pattern. A$^{+e}$A is shown in Fig.\ref{contacts}c. Then the right end of the chain makes one circle around the horizontal segment and forms A$^{+e}$B$^{+e}$AB from Fig.\ref{x}. Another similar circle around the horizontal segment will lead to A$^{+e}$B$^{+e}$C$^{+e}$ABC. A simple calculation shows that this chain indeed has the Alexander polynomial we expected. It would be interesting to investigate further on the relationship between Alexander polynomials and s-contacts (the systematic representation in Fig.\ref{loops} might be helpful), but it is beyond the scope of the present paper. Here we only hypothesize that such a relation exists. One should, however, note that the chains in Fig.\ref{x} have a different number of crossings, correspond to prime knots which also have another number of crossings; yet they all have Alexander polynomials of the same degree. We attribute it to the pattern we outlined.

\begin{figure}[ht]
\centering
\includegraphics[angle=0, width=0.6\textwidth]{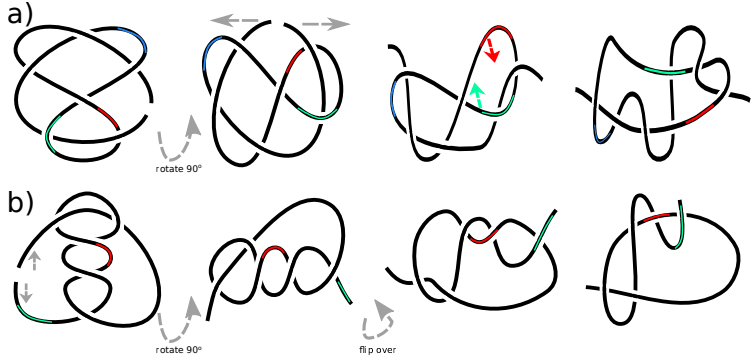}
\caption{Deformations of prime knots from shapes in Fig.\ref{prime_knots} to shapes in Figs.\ref{x} and \ref{cp}. To simplify tracing, some segments are marked with colored stripes. a) $6_2$ knot corresponds to A$^{+e}$B$^{+o}$AB. b) $5_2$ knot corresponds to (A$^{+e}$B$^{+e}$)AB (not to be confused with A$^{+e}$B$^{+e}$AB) which is not concerted.}
\label{x_to_prime}
\end{figure}

Despite having the same Alexander polynomials, the configurations from Fig.\ref{x} do not visually resemble the prime knots from Fig.\ref{prime_knots}. This occurs because of the way the loops intertwine in cross configurations. How can we be extra sure that they are actually the same, i.e., that they can be continuously deformed into each other? One way is to deform them both into the loop representation from Fig.\ref{loops}. However, such manipulations require many-step, major deformations which are hard to follow and, quite frankly, tedious to draw. More importantly, this would have no practical use. Indeed, we aim at describing proteins and other linear macromolecules. The 4 s-contacts ($+3_1$, $-3_1$, $+4_1$, $-4_1$) are to be found and identified automatically by a computer, not by a naked eye. Fig.\ref{loops} is good for visualizing the formalism and tying simple knots, but nothing more. Yet, it does look suspicious that s-contacts are not obvious in prime knots. Fig.\ref{x_to_prime}a shows the equivalence of $6_2$ knot from Fig.\ref{prime_knots} and A$^{+e}$B$^{+o}$AB form Fig.\ref{x}. Surprisingly, it requires only a minor deformation. $5_1$ and $6_3$ can be treated similarly. $5_2$ and $6_1$ will be discussed in the next section.

\section{Concerted contacts}

Concerted contacts first appeared in the context of {\bf intra-chain} contacts which are not due to entanglement. Unlike s-contacts, intra-chain contacts have specific locations of their contact sites. What happens when two contact sites are too close to each other so that they cannot be distinguished, as often the case in experiments? Such contact sites are placed inside parentheses and are possible in two variants: concerted series A(AB)B, which resembles A(BA)B, and concerted parallel (AB)BA, which resembles (BA)BA. When it comes to entanglement, contact sites of s-contacts are not specific and can move along the chain. However, some form of concerted structures still appears.

\subsection{C configuration}

Each s-contact consists of two loops. Two s-contacts cannot have more than 4 loops. The case when two s-contacts have 4 loops is considered in the previous section. However, two s-contacts can have three loops, see Fig.\ref{loops} where the middle loop is associated with both s-contacts. This ambiguity in number of loops arises from the fact that loops are not invariant of entanglement, while s-contacts are invariant.

Two loops of each s-contact must have the same chirality, otherwise it would not be an s-contact because the structure will untie. With three loops it is problematic because the loop in the middle is shared by two s-contacts. Three loops give rise to $2^3=8$ configurations. Two of them, where all the loops have the same chirality, i.e., $+++$ and $---$, are identical to two s-contacts in cross, namely A$^{+e}$B$^{+e}$AB and A$^{-e}$B$^{-e}$AB. It is expectable because in these configurations there is no chirality conflict between the s-contacts via the shared loop. The other 6 configurations are $--+$, $+--$, $++-$, $-++$, $-+-$, $+-+$. Note that the first pair and the second pair possess the left/right symmetry, hence one from each pair can be omitted. Among $+--$, $++-$, $-+-$, $+-+$, two pairs are symmetric with respect to chirality ($+\leftrightarrow-$), which leads to flipping all crossings. So, we have only two configurations left to consider, $+--$ and $+-+$. Each of them leads to 4 cases, depending on how the two pairs of loops are hooked to each other. $+-+$ either unties or coincides with A$^{+e}$A or A$^{+o}$A. So, all together, $+-+$ and $-+-$ give rise to all s-contacts.

\begin{figure}[ht]
\centering
\includegraphics[angle=0, width=0.6\textwidth]{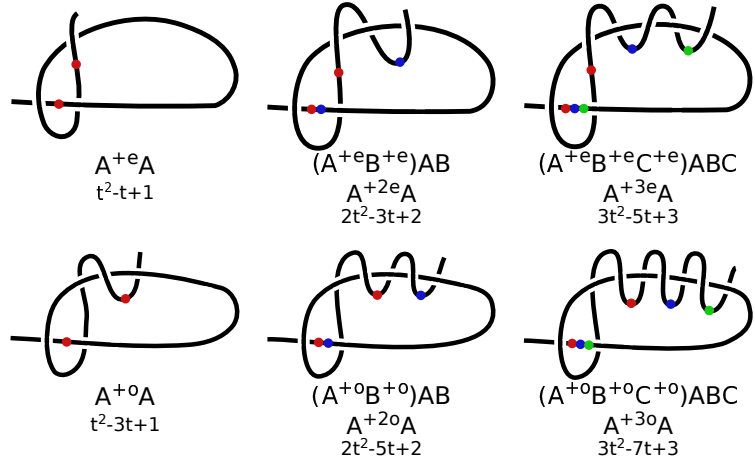}
\caption{Concerted s-contacts. Contact sites are depicted with colored balls. String notation and Alexander polynomials are provided.}
\label{cp}
\end{figure}

The last remaining configuration, $+--$, leads to (A$^{+e}$B$^{+e}$)AB and (A$^{+o}$B$^{+o}$)AB, shown in Fig.\ref{cp}. In these configurations, contacts A and B share a contact site and will both untie simultaneously if this loop is unhooked. In terms of Fig.\ref{loops}, they share a loop and will untie if the loop is undone. As stated in the previous section, a transition between the ``flat'' representation in Fig.\ref{loops} and the representation in Fig.\ref{cp} typically requires major deformations and is hard to follow. However, one can see some resemblance. Let us consider the first line in Fig.\ref{cp}. Contact A in A$^{+e}$A has two loops. Then, the right end of the chain goes around and forms contact B which shares one loop with contact A (where red and blue balls coincide). So, two loops from contact A plus two loops from contact B, accounting for one shared loop gives 2+2-1=3 loops. The left-most loop is shared and, if untied by flipping the left-most crossing, the both s-contacts disappear. The right-most loop belongs to and unties only contact B. The middle loop, which belongs only to contact A, cannot be untied all along since it is in the middle. Hence, one cannot untie only contact A, which can be seen from the illustration as well. When contact C is formed, there are 4 loops and the left-most loop is shared by all the s-contacts. Based on the analogy with intra-chain (i.e., non-entanglement) contacts, if contacts share contact sites, they are called concerted. The configurations from Fig.\ref{cp} are similar to a concerted parallel configuration of intra-chain contacts because contact A in somewhat inside of contact B. There is no concerted cross configurations for intra-chain contacts. However, the situation is different in case of entanglement, because this concerted configuration is closer to a cross rather than a parallel configuration, because the loops of s-contacts in cross cannot be separated, see Fig.\ref{loops}. This configuration is called concerted, or C, and is written as a cross configuration, i.e., (AB)AB, not (BA)AB. Also note that ABAB and (AB)AB are two different configurations, X and C correspondently. To some extent, C configuration might seem as a specific case of X, and some of their properties are indeed similar, but not identical. What makes C resemble X rather than P and S, is that C cannot be turned into another configuration (unlike P and S). 

As clear from Fig.\ref{cp}, a C configuration is possible only for the same kind of s-contacts, e.g., (A$^{+e}$B$^{-e}$)AB is impossible. Indeed, after passing through the large loop and forming contact A$^{+e}$A, the chain should pass the same large loop in the opposite direction in order to form a contact with a different chirality, i.e., contact B$^{-e}$B. But it forms a slip-knot, which is not an s-contact and will be considered in the next sub-section. Hence, the notation can be simplified: (A$^{+e}$B$^{+e}$)AB $\rightarrow$ A$^{+2e}$A, where the digit signifies the number of s-contacts. Note that C configuration is not symmetric. If the rope is read from another end, the full string notation should be used, e.g., A$^{+e}$B$^{+e}$(AB). This simplified notation is used in Fig.\ref{prime_knots} because, unlike chains, mathematical knots are closed structures and need to be cut to become chains. No matter where a prime knot is cut, the resulting linear chain is always the same. Using a simplified notation is desirable because it stresses that concerted s-contacts are, in essence, only one contact because it can be untied by one move unhooking only one loop. Alexander polynomials of concerted contacts, Fig.\ref{cp}, are degree two, which corresponds to one s-contacts. The factor at $t^2$ shows the number of s-contacts in the concerted contact, e.g., $3t^2-5t+3$, which is A$^{3o}$A, consists of $3$ s-contacts. This is in line with the conjecture made in the previous section arguing that the Alexander polynomial degree scales with the number of s-contacts for all SP, X and C configurations. SP is similar to connected sum and the corresponding scaling of Alexander polynomial is well-known in knot theory. C corresponds to a special class of knots called twist knots, for which the scaling was proved as well. X, as far as we know, does not have a direct counterpart in knot theory, and so the scaling conjunction for X has not been proven (or even considered) yet. 

Concerted contacts are closely related to each other. If we twist a rope once, we form a loop. By threading the rope through this loop, we get A$^e$A. If we twist a rope twice and thread it, we get A$^o$A, see Fig.\ref{cp}. If we twist a rope three times and thread it, we get A$^{2e}$A. Twisting four times gives A$^{2o}$A, etc. This is the twisted hairpin mechanism\cite{flapan,taylor} which leads to a formation of twist knots. All the knots found in proteins can be tied by this mechanism. In other words, concerted and only concerted contacts have been found in proteins. Here we would like to point out that C configuration does not arise as a result of some artificial mechanism of twisting the rope, but it comes out from the bottom-up consideration of possible arrangements of contact sites of s-contacts, which is more fundamental and general. The swirling part of concerted knots can be seen in Fig.\ref{prime_knots}, especially for $5_2$ and $6_1$. However, their s-contacts are still not so easy to spot. Fig.\ref{x_to_prime}b demonstrates the series of deformations needed to transition from Fig.\ref{prime_knots} to Fig.\ref{cp} for $5_2$. $6_1$ can be treated similarly. The location of the cut is irrelevant. The cuts in Fig.\ref{x_to_prime} are chosen so that the smallest deformation is required.

\subsection{Slip-knots}

\begin{figure}[ht]
\centering
\includegraphics[angle=0, width=0.6\textwidth]{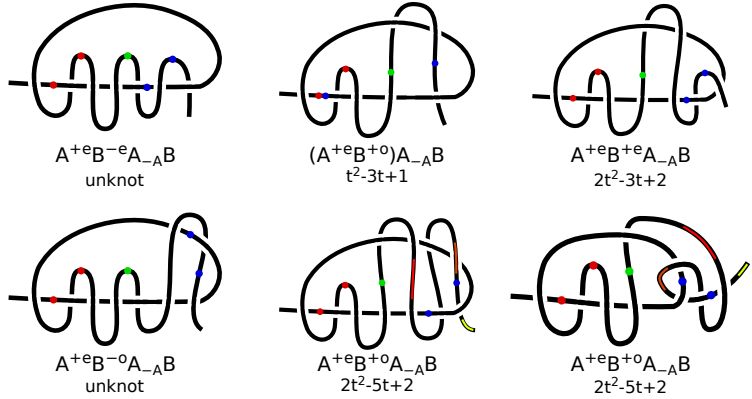}
\caption{Configurations with subscripts. Contact sites are shown as red (contact A) and blue (contact B) balls. The subscript is shown as a green ball. String notation and Alexander polynomials are provided. If the rope is followed from left to right, then the positive direction of contact A is up, towards the reader.}
\label{cs}
\end{figure}

As we stated above, (A$^{+e}$B$^{-e}$)AB is impossible. What if we try to tie it anyway? After all, we are interested in considering all possible configurations. Fig.\ref{cs} shows the treatment of even contacts, i.e., A$^e$A. Odd contacts, i.e., A$^o$A, can be treated similarly. In order to form a negative contact along with a positive contact, one has to reverse the direction of the chain. In Fig.\ref{cs} it is marked with a green ball which is located where the chain passes through the s-contact. The left part of the rope up to the green ball is the same in all configurations. Let us consider it. Contact A (red balls) consists of two loops. The second red ball shows where the rope passes through the large loop, closing (or ``fixing'') contact A. Then, the rope passes through the large loop again (green ball), so potentially it might be a sign of another s-contact. However, in this case no s-contact is formed because it does not match any structure from Fig.\ref{contacts}. In fact, we end up with the structure similar to the first or the last ones in Fig.\ref{contacts}c but with the middle crossing flipped. So, we have an event worth noticing (passing through the large loop) but there is no s-contact. In the string notation, this event is shown as a subscript, where the sign indicates the direction in which the rope passes the s-contact: ``$+$'' if it coincides with the positive direction of the s-contact, ``$-$'' overwise. Such a subscript forms a loop, but this loop is not fixed, so is does not form an s-contact. To some extent, subscripts can be viewed as a half of an s-contact. In other words, if the rope passes through a loop and fixes it, it creates an s-contact; overwise it is unknot. If the rope passes through an s-contact and fixes it, it creates another s-contact; overwise it creates a subscript. Subscripts appear only when the rope passes through an s-contact. For example, in Fig.\ref{x} the rope passes through various loops, especially in configurations with different symmetries of contacts. But it does not create subscripts because those loops are not s-contacts.

S-contacts are stable, i.e., cannot disappear or be untied, while subscripts are not stable. In any configuration shown in Fig.\ref{cs}, the rope segment between the second red and green balls forms a slip-knot which can be untied by pushing it back through contact A (red balls), which would eliminate contact A and the corresponding subscript. In terms of knot theory, this transition is a sequence of Reidemeister moves. Hence, knot theory does not see subscripts and slip-knots. Why does circuit topology ignore unstable loops but consider unstable subscripts? It is because one has to find a balance between unstable and irrelevant. Loops are very flexible structures which can appear and disappear easily. Subscripts, even though they can disappear, are more stable than loops and are observed in proteins as slip-knots. Whether a subscript is ``metastable'' or not, depends on the physical properties of the rope and the size of the corresponding s-contact. A related issue in the context of stability is that, while forming a subscript, we do not specify where exactly the s-contact is pierced by the rope. To some extent, here the s-contact is considered as a loop, not as two loops. So far we deem it as sufficient because here the s-contact works only as a restriction of motion of the rope. What part of the s-contact exactly restricts the motion is less relevant.

Fig.\ref{cs} shows two equivalent representations of the same rope configuration, A$^{+e}$B$^{+o}$A$_{-A}$B, i.e., they can be deformed into each other. The deformation is outlined in the figure by colored stripes and should be performed in 3D because these illustrations are not projections, but 3D structures. Let us consider contact B (blue balls) in both of them. In the chain on the left, contact B resembles that s-contact in A$^{+e}$B$^{+o}$AB from Fig.\ref{x}. In the chain on the right, contact B resembles that A$^{+o}$A from Fig.\ref{contacts}d. Note that the both chains are visually different but have the same Alexander polynomial, same string notation, correspond to the same knot, and can be deformed into each other. In terms of circuit topology, these two chains are equivalent. If these chains were an actual molecule, which configuration would the molecule attain? We do not know because it depends on the physical properties of the molecule.

The first column in Fig.\ref{cs} shows the rope going downwards after the subscript, i.e., after the green ball. In these configurations the chirality of the s-contacts are different and they are unknot, i.e., will untie if the chain ends are pulled. The other configurations show the rope going upwards, which leads to the same chirality of s-contacts. These configurations cannot be untied. So, consider the chirality pattern in circuit topology. If two loops have different chirality, they are unknot. If two loops have the same chirality, they form a stable s-contact. One level up: two s-contacts with the same chirality can form a stable concerted contact; while two s-contact with different chirality can form a slip-knot which can be untied. Untied configurations with subscripts with the same chirality of s-contacts lead to concerted configurations which must have the same chirality. The symmetry between concerted configurations and configurations with subscripts goes further. In this paper we first derived all possible s-contacts and then considered all possible stable configurations of them. In Fig.\ref{cp} we found that some combinations of s-contacts are impossible to be concerted. However, in Fig.\ref{cs} we see that these impossible configurations can be described by subscripts. So, concerted configurations and configurations with subscripts are complimentary to each other. Also, laying out the formalism, we try to build an analogy between intra-chain (non-entanglement) contact and s-contact which describe entanglement. Since a subscript is similar to a half of an s-contacts, contact A (red ball) and the subscript (green ball) can be vaguely considered as being in series and sharing a contact site via the slip-knot. Hence, configurations with subscripts are analogous to concerted series configurations of intra-chain contacts. In other words, we found that entanglement and intra-chain contacts have the same set of configurations in term of circuit topology, namely series, parallel, cross configurations for both intra-chain contacts and s-contacts, concerted parallel for intra-chain contacts vs. concerted s-contacts, and concerted series for intra-chain contacts vs. subscript configurations for s-contacts. This analogy matters for a completeness of circuit topology description of entanglement. Note that in case of intra-chain contacts, concerted contacts are more similar to SP configurations, while in case of entanglement, concerted contacts are more similar to X configuration.

As proclaimed above, we want circuit topology to be able to describe molecular operations easily and intuitively. It can be achieved by means of string notation. Let us consider (A$^{+e}$B)$^{+o}$A$_{-A}$B from Fig.\ref{cs} because it has both kinds of concerted contacts. Let us cut it at the loop of the slip-knot, i.e., between red and green balls. In string notation it looks like (A$^{+e}$B)$^{+o}$A\,$\mid$\,$_{-A}$B. The letter B in the left part occurs only once, so it cannot be fixed, hence it disappears. The left part turns into A$^{+e}$A. This reflects the well-known fact that a cut slip-knot becomes a trefoil. The right part, $_{-A}$B, has no pairs of letters, so it is an unknot. One can also see how the two parts are intertwined, since one part can be trapped inside of another part. The number of such connections equals the number of broken contacts in cross plus the number of subscripts. In our case there are no contacts in cross and there is only one subscript. Hence, the right part of the original chain passes once through contact A of the left part. Let us do the same manipulation with A$^{+e}$B$^{-e}$A$_{-A}$B from Fig.\ref{cs}. We get A$^{+e}$B$^{-e}$A\,$\mid$\,$_{-A}$B. The left part is a trefoil; the right part is an unknot. Due to the subscript and broken contact B, the right part passes twice througt the left part. The left part passes only once through the right part (due to contact B). One should be careful with this last statement because the right part has no s-contacts, so one cannot strictly pass though it. However, the right part swirls around the left part, forming a kind of a tunnel which the left part passes through. It can be visualized in Fig.\ref{cs}.

\section{Conclusion}

Both knot theory and circuit topology aim to describe entanglement. Knot theory considers any entangled chain as a connected sum of prime knots\cite{prime}. Prime knots cannot be divided; they are undecomposable. Circuit topology splits any entangled chains (including prime knots) into basic structural units called s-contacts, and lists simple rules how s-contacts can be put together. These rules can be considered as binary operations defined on s-contacts. There are 3 main operations (SPX) which put two s-contacts in series (S), in parallel (P), or in cross (X); and two auxiliary operations, which make s-contacts concerted (C), or add subscripts (Sub). X, C and Sub cannot be changed, while S and P can be turned into each other as long as no other operation is present. It gives rise to an interesting algebra of these operations. We found in Section 2 that contacts as a whole can be dragged along the string, which explains the transition between S and P. What would happen if other operations were present? Let us consider ACABCB. Contacts A and B are in series, but they can never become in parallel because they cannot be dragged along the string. The dragging is blocked by contact C. Indeed, if we want to drag contact A, we would also have to drag everything locked between the letters ``A'', i.e., the letter ``C''. But it is not the whole contact, hence such a drag is forbidden. In our previous work \cite{ct} we introduced the notion of circuits. A circuit is a segment of a string which consists only of pairs of letters and subscripts of the same letters. In other words, a circuit can be isolated from other contacts. By ``isolated'' we mean ``can be put in series''. Circuits can be dragged along the string. Obviously, circuits can consist of several circuits, e.g. AABCBC consists of AA and BCBC. If a circuit does not contain smaller circuits, then it is undecomposable and hence correspond to a prime knot. It would be interesting to further investigate this algebra and the detailed construction of prime knots out of circuits. For example, we said that SP looks similar to a connected sum. Why? It is because the circuit AABB consists of smaller circuits, namely AA and BB, and hence AABB is not a prime knot, which implies that it must be a connected sum of prime knots. In principle, coding entanglement as a string of letters offers an advantage of being able to apply combinatorial analysis (even before considering the algebra of circuit topology operations). In this paper, we employed it in a very mild proportion in order to count the number and kind of possible s-contacts (A$^{+e}$A, A$^{+o}$A, A$^{-o}$A, A$^{-e}$A), see Fig.\ref{contacts}, and all possible configurations of pairs of s-contacts, Fig.\ref{loops}. Indeed, two s-contacts cannot have more then 2+2=4 loops, and we considered all configurations consisted of 2, 3, and 4 loops. This pair-wise consideration is sufficient to code entanglement, i.e., to specify the unique string corresponding to the chain, but, as we just saw by ACABCB, the dynamics of the chain, i.e., the mobility of s-contacts can be affected by other contacts, so that larger scale structures such as circuits have to be considered.

So far we have considered only chains consisted of a small number of s-contacts. It might be sufficient when it comes to molecular engineering since all the knots so far found in proteins consist only of 1 or 2 s-contacts, Fig.\ref{prime_knots}. While listing these knots, we did not specify their chirality because it does not lead to any topological distinction but only flips all the crossings in the knot. However, sometimes in the literature their chirality is reported \cite{flapan}, namely the knots in proteins are A$^{+e}$A, A$^{-e}$A, A$^o$A, A$^{-2e}$A, A$^{+2o}$A, which are $+3_1$, $-3_1$, $4_1$, $-5_2$, $+6_1$. As said above, A$^o$A or $4_1$ is achiral, hence one cannot specify its sign as long as it is not in cross with other s-contacts. So, this list contains all single s-contacts (A$^{+e}$A, A$^{-e}$A, A$^o$A) and two s-contacts concerted (A$^{-2e}$A, A$^{+2o}$A). Why does this list not contain A$^{+2e}$A, A$^{-2o}$A? It has been agreed upon \cite{taylor} that topology cannot answer this question because it is related to the chemical structure of a protein chain. Also, it might be the case that these two configurations do exist, just have not been found yet. All these 5 found knots consist of concerted s-contacts only (single s-contacts are considered as a limiting case of concerted). The physical reason behind this is still unknown and lies beyond pure topology and the scope of this paper, though some speculations can be made. In order to tie a concerted structure, one has to thread a chain through a loop only once; whereas other configurations (knots) require two events of threading, thereby making them more complicated to tie. Another reason might be related to the 3D shape of the chain. To tie a concerted structure, one has to twist the chain a few times in order to form the spiral-like shape, see Fig.\ref{cp}. Such a shape might be natural for proteins and induce less stress on the chain than other shapes. In other words, the twisting motion can be done automatically by the chain itself in order to attain the preferable spiral-like shape. Circuit topology might be a convenient approach to work with such problems because it can be naturally generalized to account for relevant physical properties. Indeed, circuit topology differentiates between stable configurations (s-contacts), meta-stable configurations (subscripts, i.e., slip-knots), and not-stable configurations (single loops). Each kind of s-contact possesses its own energy; and a transition between s-contacts requires some energy (maybe in a form of entropy penalty). By building up knots out of s-contacts, one can analytically estimate the energetical complexity of various transitions.

\begin{figure}[ht]
\centering
\includegraphics[angle=0, width=0.6\textwidth]{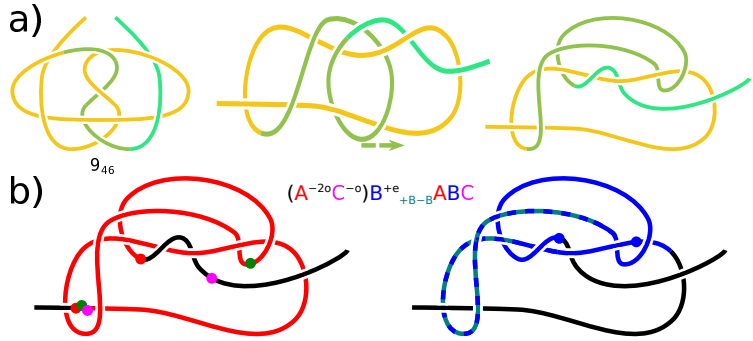}
\caption{Knot $9_{46}$. a) A 3D deformation of the knot. Segments are colored to simplify the tracing. b) S-contacts are color-coded and shown separately. The loop creating subscripts is shown in dash.}
\label{knot46}
\end{figure}

All the illustrations shown so far came from the pursue to consider all possible configurations consisted of 1 and 2 s-contacts. It is a bottom-up approach when we combine the ``basic units'' and see which knots we end up with. Let us now go in the opposite direction. We will consider a fairly complicated knot and break it down to s-contacts. As mentioned above, it is a tedious procedure which should be done by a computer, not by a naked eye. On the other hand, it helps to visualize and appreciate how s-contacts work in real life. We chose to consider knot $9_{46}$ because it has the same Alexander polynomial as knot $6_1$ from Fig.\ref{prime_knots}. In this paper, we use Alexander polynomials only to distinguish between knots while developing our approach. Alexander polynomials work very well, but fail is some rare cases. Let us see if our circuit topology can catch the different between $6_1$ and $9_{46}$. Fig.\ref{knot46}a shows a sequence of moves to deform $9_{46}$ to a more eye-friendly representation with one large loop. All the moves are in 3D. The string notation is (A$^{-2o}$C$^{-o}$)B$^{+e}_{+B-B}$ABC. Fig.\ref{knot46}b color-codes the s-contacts. Every rope segment trapped by a loop restricting its motion, gives rise to an s-contact site or to a subscript. Notice the use of the simplified notation for C configuration and the treatment of the subscripts originated from the loop passing through contact B (marked in dash). S-contacts B and C are in parallel. However, they cannot be deformed to be in series because their cross relation to contact A does not allow any transformation. So, circuit topology clearly differentiates between $6_1$ and $9_{46}$. Note that $9_{46}$ contains the same s-contact, A$^{2o}$A, which comprises $6_1$. 
Also note that $6_1$ and $9_{46}$ contain a different number of s-contacts, hence the scaling of Alexander polynomial with the number of s-contacts does not hold in this case. The main reason for this is the presence of subscripts which are not a part of knot theory (see Fig.\ref{cs} where the pattern is broken as well). Whether chains with mixed operations (SP and X and C) follow the same scaling is unclear.

We hope we have demonstrated how circuit topology can be used to describe simple knots consisted of just a few s-contacts. But how many are ``just a few'' in practice? This chain we considered, ACABCB, if s-contacts are assigned symmetry and chirality, leads to $4^3=64$ configurations. Half of these configurations are chirality symmetric (i.e. all crossings flipped; left-handright-hand symmetry). Also, $4^2=16$ configurations are leftright symmetric (i.e. the kinds of s-contacts A and B coincide). Half of these configurations are chirality symmetric as well and we do not want to count them twice. So, $64-32-16+8=24$ prime knots are offered by just one string consisted of 3 s-contacts. There are other strings made of 3 s-contacts. The number of different configurations grows fast with the number of s-contacts. Hence, all practically manageable chains involving reasonably complex prime knots are made of 3 or maximum 4 s-contacts (and subscripts). We believe it can be useful for engineering new molecular chains which can be compiled from a small set of these basic structural units.

\bibliography{rsc} 

\providecommand*{\mcitethebibliography}{\thebibliography}
\csname @ifundefined\endcsname{endmcitethebibliography}
{\let\endmcitethebibliography\endthebibliography}{}
\begin{mcitethebibliography}{13}
\providecommand*{\natexlab}[1]{#1}
\providecommand*{\mciteSetBstSublistMode}[1]{}
\providecommand*{\mciteSetBstMaxWidthForm}[2]{}
\providecommand*{\mciteBstWouldAddEndPuncttrue}
  {\def\EndOfBibitem{\unskip.}}
\providecommand*{\mciteBstWouldAddEndPunctfalse}
  {\let\EndOfBibitem\relax}
\providecommand*{\mciteSetBstMidEndSepPunct}[3]{}
\providecommand*{\mciteSetBstSublistLabelBeginEnd}[3]{}
\providecommand*{\EndOfBibitem}{}
\mciteSetBstSublistMode{f}
\mciteSetBstMaxWidthForm{subitem}
{(\emph{\alph{mcitesubitemcount}})}
\mciteSetBstSublistLabelBeginEnd{\mcitemaxwidthsubitemform\space}
{\relax}{\relax}

\bibitem[Lim and Jackson(2015)]{lim}
N.~C.~H. Lim and S.~E. Jackson, \emph{Journal of Physics: Condensed Matter},
  2015, \textbf{27}, 354101\relax
\mciteBstWouldAddEndPuncttrue
\mciteSetBstMidEndSepPunct{\mcitedefaultmidpunct}
{\mcitedefaultendpunct}{\mcitedefaultseppunct}\relax
\EndOfBibitem
\bibitem[Su{\l}kowska \emph{et~al.}(2012)Su{\l}kowska, Rawdon, Millett,
  Onuchic, and Stasiak]{stasiak}
J.~I. Su{\l}kowska, E.~J. Rawdon, K.~C. Millett, J.~N. Onuchic and A.~Stasiak,
  \emph{Proceedings of the National Academy of Sciences}, 2012, \textbf{109},
  E1715--E1723\relax
\mciteBstWouldAddEndPuncttrue
\mciteSetBstMidEndSepPunct{\mcitedefaultmidpunct}
{\mcitedefaultendpunct}{\mcitedefaultseppunct}\relax
\EndOfBibitem
\bibitem[Kuhlman and Bradley(2019)]{biofunction}
B.~Kuhlman and P.~Bradley, \emph{Nature Reviews Molecular Cell Biology}, 2019,
  \textbf{20}, 681--697\relax
\mciteBstWouldAddEndPuncttrue
\mciteSetBstMidEndSepPunct{\mcitedefaultmidpunct}
{\mcitedefaultendpunct}{\mcitedefaultseppunct}\relax
\EndOfBibitem
\bibitem[Hartl(2017)]{diseases}
F.~U. Hartl, \emph{Annual Review of Biochemistry}, 2017, \textbf{86},
  21--26\relax
\mciteBstWouldAddEndPuncttrue
\mciteSetBstMidEndSepPunct{\mcitedefaultmidpunct}
{\mcitedefaultendpunct}{\mcitedefaultseppunct}\relax
\EndOfBibitem
\bibitem[Adams(2004)]{adams}
C.~C. Adams, \emph{The Knot Book}, American Mathematical Society, 2004\relax
\mciteBstWouldAddEndPuncttrue
\mciteSetBstMidEndSepPunct{\mcitedefaultmidpunct}
{\mcitedefaultendpunct}{\mcitedefaultseppunct}\relax
\EndOfBibitem
\bibitem[Golovnev and Mashaghi(2020)]{ct}
A.~Golovnev and A.~Mashaghi, \emph{iScience}, 2020, \textbf{23}, 101492\relax
\mciteBstWouldAddEndPuncttrue
\mciteSetBstMidEndSepPunct{\mcitedefaultmidpunct}
{\mcitedefaultendpunct}{\mcitedefaultseppunct}\relax
\EndOfBibitem
\bibitem[Flapan \emph{et~al.}(2019)Flapan, He, and Wong]{flapan}
E.~Flapan, A.~He and H.~Wong, \emph{Proceedings of the National Academy of
  Sciences}, 2019, \textbf{116}, 9360--9369\relax
\mciteBstWouldAddEndPuncttrue
\mciteSetBstMidEndSepPunct{\mcitedefaultmidpunct}
{\mcitedefaultendpunct}{\mcitedefaultseppunct}\relax
\EndOfBibitem
\bibitem[Taylor(2007)]{taylor}
W.~R. Taylor, \emph{Computational biology and chemistry}, 2007, \textbf{31},
  151--162\relax
\mciteBstWouldAddEndPuncttrue
\mciteSetBstMidEndSepPunct{\mcitedefaultmidpunct}
{\mcitedefaultendpunct}{\mcitedefaultseppunct}\relax
\EndOfBibitem
\bibitem[Mashaghi \emph{et~al.}(2014)Mashaghi, van Wijk, and Tans]{2014}
A.~Mashaghi, R.~J. van Wijk and S.~J. Tans, \emph{Structure}, 2014,
  \textbf{22}, 1227--1237\relax
\mciteBstWouldAddEndPuncttrue
\mciteSetBstMidEndSepPunct{\mcitedefaultmidpunct}
{\mcitedefaultendpunct}{\mcitedefaultseppunct}\relax
\EndOfBibitem
\bibitem[Scalvini \emph{et~al.}(2020)Scalvini, Sheikhhassani, Woodard, Aupič,
  Dame, Jerala, and Mashaghi]{trends}
B.~Scalvini, V.~Sheikhhassani, J.~Woodard, J.~Aupič, R.~T. Dame, R.~Jerala and
  A.~Mashaghi, \emph{Trends in Chemistry}, 2020, \textbf{2}, 609--622\relax
\mciteBstWouldAddEndPuncttrue
\mciteSetBstMidEndSepPunct{\mcitedefaultmidpunct}
{\mcitedefaultendpunct}{\mcitedefaultseppunct}\relax
\EndOfBibitem
\bibitem[Mashaghi(2021)]{notice}
A.~Mashaghi, \emph{Notices of the American Mathematical Society}, 2021,
  \textbf{68}, 420--423\relax
\mciteBstWouldAddEndPuncttrue
\mciteSetBstMidEndSepPunct{\mcitedefaultmidpunct}
{\mcitedefaultendpunct}{\mcitedefaultseppunct}\relax
\EndOfBibitem
\bibitem[Richard \emph{et~al.}(2017)Richard, Stalter, Siebert, Rieger, Trefz,
  and Virnau]{sptran}
D.~Richard, S.~Stalter, J.~T. Siebert, F.~Rieger, B.~Trefz and P.~Virnau,
  \emph{Polymers}, 2017, \textbf{9}, 55\relax
\mciteBstWouldAddEndPuncttrue
\mciteSetBstMidEndSepPunct{\mcitedefaultmidpunct}
{\mcitedefaultendpunct}{\mcitedefaultseppunct}\relax
\EndOfBibitem
\bibitem[Schubert(1949)]{prime}
H.~Schubert, \emph{Die eindeutige Zerlegbarkeit eines Knotens in Primknoten},
  Springer-Verlag Berlin Heidelberg, 1949, pp. 57--104\relax
\mciteBstWouldAddEndPuncttrue
\mciteSetBstMidEndSepPunct{\mcitedefaultmidpunct}
{\mcitedefaultendpunct}{\mcitedefaultseppunct}\relax
\EndOfBibitem
\end{mcitethebibliography}
\bibliographystyle{rsc} 

\end{document}